\documentclass[a4paper]{article}
\usepackage{graphics}

\newcommand{\tensor}{\otimes}
\newcommand{\bra}[1]{\langle #1\vert}
\newcommand{\ket}[1]{|#1\rangle}
\newcommand{\braket}[2]{\langle #1\vert#2\rangle}

\newcommand{\unit}{\mathbbm{1}}

\usepackage{amsfonts}
\usepackage{bbm}
 
\begin{document}


\title{On the Probabilistic Compatibility of Special Relativity and
Quantum Mechanics}\author{Thomas Marlow\thanks{email:
pmxtm@nottingham.ac.uk}\\ \emph{School of Mathematical
Sciences, University of Nottingham,}\\\emph{UK, NG7 2RD}}

\maketitle

\begin{abstract}

In this paper we introduce the three main notions of probability used
by physicists and discuss how these are to be used when invoking
spacelike separated observers in a relativistic format. We discuss a
standard EPRB experiment and concentrate upon problems of the
interpretation of probabilities. We promote a particularly
conservative interpretation of this experiment (which need not invoke an
objective notion of collapse) where probabilities are, tentatively,
passively Lorentz invariant. We also argue that the Heisenberg
picture is preferable in relativistic situations due to a conflict
between the Schr\"{o}dinger picture and passive Lorentz
transformations of probabilities. Throughout most of this paper we
discuss the relative frequency interpretation of probability\----as
this is most commonly used. We also introduce the logically necessary
notion of `prior-frequency' in discussing whether the \emph{choice}
by an observer can have any causal effect upon the measurement
results of another. We also critically examine the foundational use
of relative frequency in no-signalling theorems. We argue that SQT
and SR are probabilistically compatible, although we do not discuss
whether they are compatible on the level of individual events. 

\end{abstract}

\subsection*{Motivation}

In the literature it is often stated that there is some kind of
peaceful coexistence$^{(\ref{Shimon83})}$ between Special Relativity (SR) and
Standard Quantum Theory (SQT). This is
usually expressed by the statement that SQT cannot be used to send
informative signals at faster than the speed of light. Sometimes
it is also stated in stronger terms, with the probabilities of
outcomes of experiments being manifestly Lorentz invariant$^{(\ref{Peres00})}$.

With so many interpretations of SQT abound, it would seem prudent
to analyse this notion of peaceful coexistence and see which
interpretations express it most cogently. For example, often the
more realist interpretations of SQT (those that involve an objective
notion of wavefunction collapse) are said to be inconsistent with SR. In
this paper we shall take a very conservative stance; we shall
only discuss the `probabilities' of outcomes of experiments and
thus this issue shall be skirted in the main discussion. All arguments
below should be independent of any ontological stance one takes, as long
as one accepts a notion of collapse.  We shall not discuss no-collapse
pictures in this paper.

The aim of this paper is to try and find out how well our operational
notions of probability mesh with SQT and SR, and whether we can promote one
interpretation of probability over another.  We aim to analyse peaceful
coexistence in the light of these foundational issues, and show
causality to be ambiguously defined on a probabilistic level (this is
of course almost trivial but it is useful to state it pedagogically).
Recently Stapp$^{(\ref{Stapp03})}$ has attempted a proof of
nonlocality in SQT by only invoking logical relationships between
individual events. Stapp's approach has recently been criticised by
Shimony$^{(\ref{Shimon04})}$. We argue, here, only that nonlocality
cannot be proved or disproved using probabilistic concepts alone; we
do this in a way that is pedagogically useful, through the
introduction of a notion we call ``prior-frequency''.

But how are we to interpret `probability'?  There are a few
different notions of probability that physicists use, most of
which can be put somewhere within the following (surprisingly
broad) three categories:

\begin{enumerate}

\item \emph{Relative Frequency.} This is the usual notion of
`probability' discussed by physicists\----if just for the
simplicity of its formulation. Here, relative frequencies are
defined relative to infinite ensembles of `identical'
experiments. The relative frequency of an event being the
proportion of the ensemble in which this event occurred (in the
limit of infinite trials). Within this interpretation, relative
frequencies represent the unpredictability of the outcomes of a
repeat of a single experiment. It is, however, ambiguous where
the origin of unpredictability lies. It could either originate
from an inherent unpredictability of the system under
investigation, or an inherent unpredictability within the
`identical' setups used to probe that system, or both. For a
cogent description of this view see Ref. \ref{MisesBOOK}.

\item \emph{Bayesian Probability.} Here, probabilities are not
necessarily related to frequencies.  The Bayesian
view is to take the theory of probability as a ``logic of
inference''. There are a variety of Bayesian standpoints but they all
share the idea that probabilities are not objective properties of
things or ensembles, but rather, are subjective quantities that
depend on both prior-information and data. This is not to say that
probabilities are considered wholly subjective\----quite the
opposite; they are sometimes (as in the formalism of
Jaynes$^{(\ref{JaynesBOOK})}$) considered uniquely defined for all
observers with the same prior-information and data as long as they
use \emph{all} the prior-information they have. Some authors treat
Bayesian probability in a slightly more, but still not wholly,
subjective manner$^{(\ref{FinettiBOOK})}$. For a classic
introduction to the use of Bayesian probability in SQT see Fuchs' recent
treatise$^{(\ref{Fuchs02})}$. Like relative frequencies, Bayesian
probabilities are carriers of incomplete information, but in the
Bayesian case they depend, either uniquely or in part, upon the
prior-information and the experimental setup; else they would be
entirely subjective. One argument that is often given against the
Bayesian viewpoint is that of circularity; we seem to be defining
probability with a notion of `prior-probability'. Suffice to say,
this may simply be a problem of etymology, but we shall not go
into that here.

\item \emph{Propensity.} Here, probabilities are
considered objective, and represent the actual `likelihood', or
`propensity', of an event. A propensity is the tendency of a
possibility to realise itself upon repetition of the `same'
experiment. This view was proposed by Popper$^{(\ref{PopperBOOK})}$
in order to be able to apply the notion of probability to
objective single systems (where `system' here includes a notion
of the experimental setup). Although it is to be noted that,
within this interpretation, propensities are only ever calculated
via the relative frequency method.

\end{enumerate}

In the remainder of this paper we shall introduce the Bohm version of
the classic Einstein-Podolsky-Rosen (henceforth EPRB)
thought-experiment, and we will show that one cannot argue for (or
against) causal relations between events using these notions of
probability alone. In doing so we will argue for the use of the
Heisenberg picture and introduce the logically necessary notion of
``prior-frequency''.  This will also provide impetus to discuss the
foundational use of probability in no-signalling theorems.

\subsection*{Peaceful Coexistence}

If we are, tentatively, to discuss the peaceful coexistence of
two theories we must, in some sense, `trust' those two theories$^{(\ref{Rovel99})}$. If we trust SR then two spacelike separated
events cannot be causally connected. This, however, is very
different to saying that two spacelike observers cannot be
\emph{logically} connected. We shall elaborate upon this as we go.

If we trust SQT then we can predict the frequencies (or
probabilities or propensities) of outcomes of single-system
measurements. And, if we also posit some kind of collapse (by
whatever means or justification) then we can also, we believe, predict
the frequencies of outcomes of entangled subsystems, even if they are
spacelike separated.
 
To claim `peaceful coexistence' is to suggest that, trusting both
SR and SQT, neither are disobeyed in an overt manner.
A na\"{\i}ve view would suggest that this is impossible since any
change in frequencies (or probabilities or propensities) is often
said to occur instantaneously\----where `instantaneous' is usually
defined in whatever Lorentz frame we are discussing. However, it is
usually agreed that no useful information transfer between spacelike
separated observers can occur through this method. It is upon this
fact that many take SR and SQT to be loosely compatible. We say
`loosely' exactly because we might, perhaps na\"{\i}vely, take the
two theories to be incommensurable. We say `compatible' exactly
because we feel the spirit of SR remains
intact when no useful information transfer can occur at faster than
the speed of light.

\subsection*{An EPRB experiment}

In this section we shall set up a framework for the discussion of
peaceful coexistence. We will use the standard EPRB setup, using
an entangled initial spin-state for two correlated electrons
prepared at a source $S_0$, and two spacelike separated
observers $S_1$ and $S_2$ that use the Hilbert spaces
${\cal{H}}_1$ and ${\cal{H}}_2$ to describe their respective
subsystems. Any entangled state is discussed in reference to
${\cal{H}}_1 \tensor {\cal{H}}_2$. For the sake of argument, let
us frame the discussion below in terms of relative frequency; we
shall discuss the other interpretations of probability once we
have set up the problem.

\refstepcounter{figure}
\begin{figure}[h!]\label{sigma}
\unitlength=1.00mm
\begin{center}
\begin{picture}(50.00,50.00)
\bezier{200}(25.00,5.00)(13.00,26.00)(5.00,40.00)%
\bezier{200}(25.00,5.00)(42.00,35.00)(45.00,40.00)%
\bezier{200}(5.00,40.00)(25.00,45.00)(45.00,40.00)%
\bezier{200}(5.00,40.00)(26.00,35.00)(45.00,40.00)%
\put(25,5){\circle*{1}}
\put(20,30){\circle*{1}}
\put(30,30){\circle*{1}}
\put(20,1){$(S_0, t_0)$}
\put(14,25){$(S_1, t_1)$}
\put(25,33){$(S_2, t_2)$}
\end{picture}
\end{center}
\begin{center}
Figure \thefigure. Frame $\Sigma$, where $t_1=t_2$.
\end{center}

\end{figure}

The entangled initial state represents two electrons that are
released, say, in opposite directions\footnote{One could also frame this
discussion in terms of non-identical particles if one wished.}. At
a later time, $S_1$ and $S_2$ each have access to only one of
the two electrons, and they each have a choice of, say, two
different spin directions in which to measure their received
electron. For clarity, let us call the spins in these directions
$S_x$ and $S_z$ and assume that both the observers have
communicated previously so as to be discussing the same
directions.

Let us take a Lorentz frame $\Sigma$ (see Fig.\ref{sigma}.) in
which $S_1$ and $S_2$'s measurements are considered
simultaneous. In this frame we discuss the following entangled
initial state (where the notation that refers to different
subsystems is assumed obvious):

\begin{equation}
\ket{\psi}=\frac{1}{\sqrt{2}}(\ket{x+}_1\tensor\ket{x-}_2-
\ket{x-}_1\tensor\ket{x+}_2)
\end{equation}

Let us, for the moment, frame this discussion in terms of the Schr\"{o}dinger picture
and let us discuss the na\"{\i}vest form of collapse that we can think of\----where
collapse is instantaneous\footnote{It doesn't matter whether we consider whether this collapse
process is `objective' or `subjective' because the only things we ever verify in SQT
are the final probabilities\----so it is only these probabilities that we should query
philosophically.} in whichever frame of reference we are presently discussing.   We
consider $S_1$ and, without loss of generality\footnote{This step is perhaps dubious
from a probabilistic standpoint due to the infinite nature of ensembles, and we
reserve the right to query it later.}, we look at the case where she chooses to
measure the spin in the $x$-direction and $S_2$ measures spin in the $z$-direction.
When $S_1$ measures $S_x$ she will, in a single run of the experiment, either
receive the result $S_x=+1$ or $S_x=-1$ for particle one. Since, in a single run,
she will only ever receive one or the other, let us discuss the case where she
receives $S_x=+1$ for her electron. Regardless of what we take `state' to mean
(\emph{i.e.} whether it is defined physically or with respect to ensembles), $S_1$
believes she `has' the state $\ket{x+}_1$ and that, by the entanglement property and
the simultaneity of the measurements, $S_2$ must have `received' the state
$\ket{x-}_2$. $S_2$, however, does not know what $S_1$ has received and he will not
use the state $\ket{x-}_2$ as his initial state; he will instead measure $\unit
\tensor S_z$ using the initial state $\ket{\psi}$. $S_1$ will use $\ket{x-}_2$ to
predict the relative frequency of $S_2$'s results given her's, yet $S_2$ will use
$\ket{\psi}$. Thus, $S_1$ and $S_2$ use different methods to discuss the relative
frequencies of the `same' events, and there is no \emph{a priori} reason why they
should be the same, for they are calculated by different methods; they are
\emph{not} discussing the same ensembles. $S_1$ uses a pre-selected ensemble where
she always receives the result $S_x=+1$, which is trivially different from the
ensemble $S_2$ uses. It is also trivial that these two relative frequencies could
have different limiting values. We shall discuss the possibility of $S_1$ invoking
an unselective ensemble later.

\subsection*{On the Conflict Between the Schr\"{o}dinger Picture and Passive
Lorentz Invariance of Probabilities}

We have argued above that the relative frequencies that are
predicted for an `event' by spacelike separated observers are
\emph{not} the same and, furthermore, that they need not be the
same for logical consistency as long as one accepts that
different observers are necessarily discussing different
ensembles. If you do not agree with this statement then please
continue as we shall be discussing more `objective' relative
frequencies below. But what about the frequencies predicted by
observers discussing the same measurements in a different
Lorentz frame? Here, we are discussing passive Lorentz
transformations rather than active ones. $S_1$ can ask ``if I
were the observer $S'_1$ in frame $\Sigma'$ then what
probability would I \emph{infer} for $S'_2$'s results''. $S_1$
hopes that such inferences should be the same in all
frames\----else there may exist a way to physically
differentiate frames of reference; observers in different frames
would predict different probabilities for the `same'
measurements. If the Poincar\'{e} group induces a symmetry on
the Hilbert space then a Lorentz transformation of states is
given by a unitary operator (see chapter 2 of
Ref. \ref{WeinbergBOOK}). In this case, due to the cyclic trace
property of frequency calculations, frequencies are defined to
be Lorentz invariant. But, if we wish to invoke any collapse
postulate, the situation is more complicated than this; we must
take into account any unitary evolution and collapse of states
within each frame$^{(\ref{Myrvold02})}$.
 
\refstepcounter{figure}
\begin{figure}[h!]\label{sigmaprime}
\unitlength=1.00mm
\begin{center}
\begin{picture}(50.00,50.00)
\bezier{200}(25.00,5.00)(13.00,26.00)(5.00,40.00)%
\bezier{200}(25.00,5.00)(42.00,35.00)(45.00,40.00)%
\bezier{200}(5.00,40.00)(25.00,45.00)(45.00,40.00)%
\bezier{200}(5.00,40.00)(26.00,35.00)(45.00,40.00)%
\put(25,5){\circle*{1}}
\put(20,28){\circle*{1}}
\put(30,32){\circle*{1}}
\put(20,1){$(S'_0, t'_0)$}
\put(15,24){$(S'_1, t'_1)$}
\put(24,34){$(S'_2, t'_2)$}
\end{picture}
\end{center}
\begin{center}
Figure \thefigure. Frame $\Sigma'$, where $t'_1<t'_2$.
\end{center}
\end{figure}

In frame $\Sigma'$\----where $S'_1$'s measurement occurs before
$S'_2$'s\----the na\"{\i}vest view possible is, again, to
suggest that the collapse occurs at the time of $S'_1$'s
measurement. Again let us fix that $S'_1$ measures $S_x$ and
receives, in the Schr\"{o}dinger picture, the result $S_x=+1$ at
time $t'_1$, and that, at time $t'_2$, $S'_2$ measures $S_z$.
The frequency which $S'_1$ predicts $S'_2$ to receive the result
$S_z=+1$ given that she received $S_x=+1$ is:

\begin{equation}
\vert \bra{z+} \hat{U}(t'_2-t'_1) \ket{x-} \vert^{2}
\label{Schr}
\end{equation}

Here, we have dropped the subscripts on bras and kets as we are
discussing states in ${\cal{H}}_2$
alone. The unitary operator $\hat{U}(t'_2-t'_1)$ is the
evolution\footnote{We are implicitly assuming that there are no
nonlocal interaction terms in the total hamiltonian of
${\cal{H}}_1 \tensor {\cal{H}}_2$.} operator in ${\cal{H}}_2$.
However, frequency (\ref{Schr}) is not obviously Lorentz
invariant. When discussing a particular measurement in the
Schr\"{o}dinger picture, the quantum state invoked collapses to
one of a fixed set of post-measurement states at each use of that
measurement (the eigenstates of the operator that represents that
measurement)\----presumably in whatever frame of reference we are discussing because, in
the Schr\"{o}dinger picture, the states which correspond to the outcomes of a particular
measurement are independent of the time (and spatial position) of the measurement. At
each measurement we are, in a sense, losing information about the fiducial time at which
the initial state was prepared. Also, it seems impossible to make such a setup passively
Lorentz invariant; if we change frames then states are not going to change, but the
unitary evolution between them does. This ruins any passive Lorentz invariance that we
might tentatively have. This na\"{\i}ve notion of ``simultaneous collapse in all frames''
is therefore not explicitly Lorentz invariant when discussed in the Schr\"{o}dinger
picture.  This conflict between the Schr\"{o}dinger picture and the lorentz invariance of
probabilities is independent of any ontological significance that one assigns to the
wavefunction; we are dealing with probabilities alone and, accepting a collapse
hypothesis, frequency (\ref{Schr}) cannot be made explicitly Lorentz invariant.  We
cannot transform post-measurement states in the Schr\"{o}dinger picture and we thus cannot
maintain any Lorentz invariance of frequency (\ref{Schr}).  Thus this conflict comes
about despite the formal equivalence of the Schr\"{o}dinger and Heisenberg pictures due
to the \emph{interpretation} we must assign to post-measurement states within the
Schr\"{o}dinger picture.

Let us now try and describe the same situation in the Heisenberg
picture. In the Heisenberg picture, the possible
post-measurement states depend upon the time (and, presumably,
the spatial position) of the measurement\----hence, to put it in a
trite manner, in $\Sigma'$ we measure $S_x'$ and $S_z'$ and not $S_x$
and $S_z$.

If collapse occurs at time $t'_1$ then the resulting Heisenberg
state after measurement in subsystem one is given as
$\ket{x'+(t'_1-t'_0)}$; $S'_1$ then infers that, at time $t'_1$, the Heisenberg
state of subsystem two is the collapsed state
$\ket{x'-(t'_1-t'_0)}$. $S'_1$ can then infer, she believes,
that the relative frequency of $S'_2$'s result $S_z'=+1$ should be the
frequency given by:

\begin{equation}
\vert \braket{z'+(t'_2-t'_1)}{x'-(t'_1-t'_0)} \vert^{2}
\label{Heisfalse}
\end{equation}

This is $S'_1$'s inference about the frequencies of $S'_2$'s results
and has no bearing upon $S'_2$'s inferences\----they are space-like
separated. In relativistic situations the transformation of states
explicitly depend on the Lorentz transformation invoked; states
transform unitarily like:
\begin{equation}
\ket{\psi} \longrightarrow \hat{U}(\Lambda,a)\ket{\psi} \end{equation}
where $\Lambda$ is a Lorentz transformation, and $a$ is a translation
parameter$^{(\ref{WeinbergBOOK})}$. Both Heisenberg states in
(\ref{Heisfalse}) are defined in ${\cal{H}}_2$ (note that this
Hilbert space is not primed as we are presuming the Poincar\'{e}
group induces a symmetry on ${\cal{H}}_2$) and thus the frequency
(\ref{Heisfalse}) is explicitly Lorentz invariant. If one also wishes
for the covariance of the collapse process itself, then one might
also invoke the methods of Hellwig and Kraus$^{(\ref{HK70})}$.
 
Although the Schr\"{o}dinger and Heisenberg pictures are formally equivalent
in a non-relativistic format, the Schr\"{o}dinger picture is ambiguous when
used to invoke passive Lorentz transformations of probabilities. In the
Schr\"{o}dinger picture, post-measurement states are independent of the time
of the measurement and it is exactly these states that we wish to
transform when Lorentz transforming probabilities. It is by this that I
suggest there is a conflict between the Schr\"{o}dinger
picture and the tentative Lorentz invariance of probabilities when
discussing relativistic situations.

\subsection*{Discussion}

We have not yet discussed, however, whether any of the above is
consistent with the change in time ordering that can come about
when discussing changes in frames of reference with respect to
spacelike separated observers. If we Lorentz transform
frequency (\ref{Heisfalse}) to a frame $\Sigma''$ such that $t''_2<t''_1$
then, as we discussed above, the frequency predicted is the same
as long as the Poincar\'{e} group is a symmetry of
${\cal{H}}_2$. We must ask, however, whether the change in
temporal ordering is consistent with the \emph{interpretation}
of frequency (\ref{Heisfalse}). In $\Sigma''$, $S''_1$'s measurement
occurs after $S''_2$'s. If we wish to discuss $S''_1$'s
prediction for the frequency of $S''_2$'s result then, with the
new temporal ordering, we have to invoke a collapse
in the second subsystem that occurs \emph{in the future} of
$S''_2$'s measurement. This may seem a weird interpretation but
it is completely consistent with a post-selection of a quantum
state in an ensemble$^{(\ref{Vaidman98})}$. An observer is allowed to
reason like the following (as long as he has no information
about the previous measurement): ``What does my result suggest
about the unknown result of a previous known experiment?''.
$S''_1$ can discuss an ensemble where she post-selects her
received state, and using this ensemble she can predict the
frequency of results of $S''_2$'s previous measurement \emph{given
hers}. Although $S''_1$ will not `know' which measurement $S''_2$ made,
they can confer before the measurements take place and collaborate in
their actions\----and given that $S''_2$ doesn't mischievously
measure something else entirely then $S''_1$'s predictions for the
frequencies of $S''_2$'s results \emph{in this post-selected ensemble}
will, presumably, be correct.

\refstepcounter{figure}
\begin{figure}[h!]\label{sigmaprimeprime}
\unitlength=1.00mm
\begin{center}
\begin{picture}(50.00,50.00)
\bezier{200}(25.00,5.00)(13.00,26.00)(5.00,40.00)%
\bezier{200}(25.00,5.00)(42.00,35.00)(45.00,40.00)%
\bezier{200}(5.00,40.00)(25.00,45.00)(45.00,40.00)%
\bezier{200}(5.00,40.00)(26.00,35.00)(45.00,40.00)%
\put(25,5){\circle*{1}}
\put(20,32){\circle*{1}}
\put(30,28){\circle*{1}}
\put(20,1){$(S''_0, t''_0)$}
\put(14,34){$(S''_1, t''_1)$}
\put(24,24){$(S''_2, t''_2)$}
\end{picture}
\end{center}
\begin{center}
Figure \thefigure. Frame $\Sigma''$, where $t''_1>t''_2$.
\end{center}
\end{figure}

Thus, it seems, we can discuss the results of SQT in relativistic
situations consistently. Frequencies predicted by any given
observer can, tentatively, be passively Lorentz invariant in the
Heisenberg picture and can be interpreted consistently with
respect to different time orderings due to the ensemble nature
of relative frequencies. This
might suggest that SQT and SR don't `peacefully coexist'; they
might be more than loosely compatible. To put it gnomically, perhaps SR and
SQT do not peacefully coexist, but, rather, perhaps spacelike
separated observers do.

Peaceful coexistence is usually expressed through no-signalling
theorems, where it is proved that if we `trace out' one observer's
influence we get the same value of relative frequency as we would have
received had that observer not made their measurement at all. One problem
with such arguments is that it is not clear what ensemble, and equivalently
what relative frequency, we are discussing. No-signalling theorems (see,
for example, Ref. \ref{Peres00}) tend to be proved by invoking only local
measurements (which are automatically commutative due to the nature of the
tensor product) and they prove that the relative frequency of one
observer's (say Bob's) result given that the other observer (say Alice)
made a certain measurement (but `tracing out' Alice's results) \emph{has
the same value} as the relative frequency of Bob's result given that
Alice did not make her measurement. During these proofs no use of
spacelike separation is made, so these theorems prove that no two
subsystems can signal between each other given local measurements. These
theorems, however, need not prove anything about causality;
regardless of the fact that the value of the relative frequencies are the
same, the relative frequencies are defined by ensembles that are
\emph{counterfactually distinct}\----one is \emph{defined} in an ensemble
where Alice \emph{always} makes her
measurement (in every element of the ensemble) and the other is defined in
an ensemble where Alice \emph{never} makes her measurement.  These proofs
only prove the happy coincidence that two counterfactually distinct
ensembles happen to have the same limiting value of the relative
frequencies of a certain event. In order to prove anything about causality we must model
a choice, between these two counterfactually distinct cases, that is well-defined at a
single trial and in a small spacetime region. So, strictly speaking, no-signalling
theorems do not prove that relativistic causality is maintained; we may even require
additional `common sense' assumptions (like in Bell inequality
arguments$^{(\ref{BellBOOK})}$ or the recent argument of Stapp$^{(\ref{Stapp03})}$)
about the relations between individual events in order to even discuss causality.

No-signalling theorems suggest that two different relative
frequencies must have the same \emph{value} for local
measurements, but they say nothing about the logical distinction
of the two relative frequencies; in this sense the two relative
frequencies may `objectively' have the same value but they
cannot `objectively' be the same relative frequency. In order to
even prove no-signalling theorems one must assume that there is
no \emph{a priori} reason, regardless of causality, for the frequencies to
have the same limiting value in order to prove that they do, otherwise
the theorem would be a tautology. No two equals are the same.
Probabilities with the same value are not the same probabilities. So if
we are to call relative frequencies `objective', in the sense that all
observers agree that they are correct, we must note that their very
definition depends upon the ensembles used when invoking them; if we wish
to discuss the `actual' relative frequency we must invoke the `actual'
ensemble used.

\subsection*{`Objective' Relative Frequencies}

One argument against the `relativity' of relative frequencies might be
that relative frequencies are `objective' properties of ensemble
experiments\----could we not simply set up an experiment, let it run,
and then discuss, retrospectively, the relative frequencies of each
outcome completely objectively? In a given experiment (where no choices are made) we
have a well defined notion of relative frequency because we have fixed the
discussion to a single ensemble of results with a well defined sampling and limiting
process.  Returning to $\Sigma'$, to get the `actual' relative frequencies of `the'
experiment we simply fix what $S'_1$ and $S'_2$ measure and record the results of
each subsystem upon each repetition. Thus we define an ensemble such that in each
element of the ensemble the measurements for each observer are fixed, and are made
in every element of the ensemble, and thus we can have a well defined relative
frequency that all will agree is correct. Say $S'_1$ wishes to discuss the relative
frequency of $S'_2$'s result; what she will do, theoretically, is set up an ensemble
made up of two counterfactually distinct subensembles; one subensemble being the one
where $S'_1$ received the result $S_x'=+1$ and the other where she received
$S_x'=-1$. Let us call the frequency that $S'_1$ predicts for each of her own
results $\nu(S_x'=-1\vert\psi')$ and $\nu(S_x'=+1\vert\psi')$ and, presumably, these
are the weights she applies to each subensemble respectively. In the Heisenberg
picture (we drop the time labels however), the relative frequency she shall predict
for $S'_2$ to receive the result $S_z'=+1$ is (presuming there are no problems with
combining frequencies that are in the limit of infinite trials):

\begin{equation}
\vert \braket{z'+}{x'-} \vert^{2}\nu(S_x'=+1\vert\psi') +
\vert \braket{z'+}{x'+} \vert^{2}\nu(S_x'=-1\vert\psi')
\label{objective}
\end{equation}

There is no \emph{a priori} reason, regardless of issues about causality, why this
should give the same relative frequency that $S'_2$ predicts \emph{given that $S'_1$
didn't make a measurement}, namely $\nu(S_z'=+1\vert\psi')$. I use the term \emph{`a
priori'} in the sense that there is no \emph{logical} reason why they should be the
same; regardless of whether they have the same value, they are different relative
frequencies\----they are defined on counterfactually distinct ensembles.  If there is
no logical reason they should be the same why claim that if they are different this
is due to causal reasons?  We are discussing different ensembles so it is a tautology
that the relative frequencies could be different\----there is no \emph{a priori}
reason why causality should be invoked to explain any difference.

One should sample the results in the way that they `actually' occur before
one takes the limit\----else, since infinities rarely follow common
sense, one may get nonsense. There is nothing wrong with saying ``if the
world were different we get different frequencies of results'', and
surely we should not claim that this is a causal influence. The frequencies might be
different for reasons related to the logical distinction of the two distinct worlds
or ensembles.  And given that we cannot claim that there is a causal influence, we
also cannot claim that there is none. This obviously applies to the classical use of
relative frequencies at spacelike separation as well.

One might try and get around the above argument by discussing
\emph{either of two ensembles} (with the knowledge that these are the \emph{only}
two cases), one where $S'_1$'s measurement always takes place and one
where it always does not take place; if the two relevant relative frequencies
had different limiting values then $S'_2$ might be able to infer whether $S'_1$ made
her measurement. In this case, would we be able to infer a causal influence? Should
such signals, if they were to be possible, be considered causal or logical things?
$S'_1$ and $S'_2$ have obviously communicated previously in order to set up the
experiment, and $S'_1$ would have to decide right from the outset (but without
$S'_2$ knowing) whether to make her measurement in every element of the ensemble or
none. This the logical equivalent of a common cause.  In the hypothetical situation
where the two relevant relative frequencies did not have the same limiting values
then some would say that $S'_2$ could infer whether $S'_1$ made her measurements or
not, even though she is spacelike separated from him. Trusting that the spirit of SR
should remain intact, many of us would rather the two frequencies had the same
limiting values so that `information' cannot travel at faster than the speed of
light.

In this hypothetical world (where no-signalling theorems are false), in order to
access the relevant information $S'_2$ has had to collaborate with $S'_1$ to ensure
that she will always measure, or always not-measure, in each element of the ensemble
(but not which choice she will make). He then must experimentally determine the
relative frequency of his result and infer from that whether $S'_1$ made her
measurements or not (even though this is in principle impossible, due to the infinity
of experimental runs required, there would come a point where he is confident that he
can infer $S'_1$'s choice with a given accuracy). But could we then infer that
$S'_1$'s measurement \emph{causally} effected $S'_2$'s results? We could definitely
infer that $S'_1$'s measurement affected $S'_2$'s \emph{relative frequencies} but in
order to argue for causal relations we must model $S'_1$'s \emph{choice} of
measurement at a single trial, which is a different matter entirely. Note also that
$S'_1$ and $S'_2$ have collaborated significantly so it is not clear how we can
distinguish common causes from uncommon ones (and these causes from logical
collaborative information that both observers have).

Lets say $S'_1$ can and
does signal to $S'_2$ that she made the measurement in every element of the
ensemble. $S'_2$ knew from the outset that she would either make the measurement in
every trial or not make the measurement in every trial, for they have collaborated
so as to complete exactly this experiment, and he can infer the relative frequencies
of his results conditional on each possibility. In what way would $S'_2$ be
\emph{informed} by this signal? He cannot use this `information' to predict anything
useful happening around him locally. He would be `informed' about a spacelike
separated region, but not about his own. But by its very nature he cannot know what
is going on in a spacelike separated region without collaboration and trust. This
highlights that we are discussing \emph{inference} and not \emph{influence}\----an
inference conditional on the fact that he has collaborated with $S'_1$ and knows
exactly what she might do (and that she will do exactly the same thing in every
repeat of the experiment while he works out the relative frequencies of his
results). Again, this might be the logical equivalent of a common cause.  He can
infer what measurement she made because he collaborated with her and knew what his
relative frequencies would be in each case, and he trusts that only these two cases
are apt. Thus signalling need not obey SR in an \emph{a priori} sense as we are
discussing inferences and not material influences. Therefore no-signalling theorems
prove something slightly different to SR. In the Bayesian approach this latter
statement is trivial as probabilities are explicitly considered inferences and not
objective properties of things.

Such an experiment as above would seem a lot of
effort to go to to transmit a single bit of information; but anyway, no-signalling
theorems prove that such a method could not be used to signal such information
between spatially separate observers since the relative frequencies do have the same
value. SR only proves that light and matter cannot travel faster than \emph{c} and
thus no-signalling theorems go further and show that `information' cannot be
signalled, in the way discussed above, regardless of separation (even though it
would be a particularly inefficient way to transmit information anyway as it
requires many trials to differentiate relative frequencies). I am only arguing here
that we cannot consider such signals as strictly causal in the first place.

In order to show this, let us look a little deeper at our model of this `choice' that
$S'_1$ makes. In order for the experiment to go as planned $S'_1$ needs to `choose'
the same `choice' in each element of the ensemble. $S'_2$ must be sure that she has
done the same thing each time the experiment is repeated. In each trial $S'_1$ must
make her `choice' when she becomes spacelike separated from $S'_2$, but she must
choose the same `choice' each time. This `choice' is no choice. If we want to model
a `real' choice, where she has the possibility of choosing either option in each run
of the experiment, then we come across an inherent ambiguity in the discussion. Does
$S'_1$, say, choose to measure $S'_x$ half the time, or one percent of the time? We
have to know the relative frequency of her choice in order to set up the model. Yes,
$S'_1$'s choice of whether she will measure or not can, tentatively, be causally
separate from $S'_2$'s measurement, but we must invoke two ensembles weighted by the
frequency of $S'_1$'s choice\----we will call this a super-ensemble and the
frequency of $S'_1$'s choice a prior-frequency. This prior-frequency, if it is
well-defined, can be chosen before hand, or inferred retrospectively, but, either
way, this is trans-temporal (or rather `trans-trial') knowledge which must be taken
into account. Each such choice for the frequency of $S'_1$'s choice is a
counterfactually distinct case (in terms of ensembles). Either she measures with one
prior-frequency or she measures with another.

$S'_2$ cannot know this prior-frequency
of $S'_1$'s choice unless they decide together beforehand and in such a case $S'_2$
is not informed of anything when they complete the experiment. The case where she
either measures in every element or in no element of the ensemble cannot be
described by a single prior-frequency and this is the only case where $S'_2$ might
learn anything. No-signalling theorems prove that he cannot learn anything at all in
this case but, even if he could, we cannot prove that an event at a single trial
\emph{causes} anything to change in the spacelike separated region. To put it
gnomically, if something that is not defined at a single trial affects results then
why call it a `cause'?  Rather, this `something' is related to the logical relations
between trials and is non-trivial.  In both of the two possible ensembles $S'_1$ knew
from the outset which choice she would make in all trials. This might be the logical
analogy of a common cause: perhaps we could call it `common knowledge' in the sense
that both observers know that $S'_1$'s choice is made in all trials. The only time
we could prove anything about causality would be when $S'_1$ makes her
choice at a single trial (and at spacelike separation to $S'_2$), but such choices are
not well-defined using ensembles (which are in turn used to define relative
frequencies) because they could occur with any given prior-frequency.
 
One cannot argue for cause without good cause\----one must discuss the
connections between individual events (like Stapp$^{(\ref{Stapp03})}$ has attempted
to do). The only other option is to \emph{define} a notion of causality using
probabilistic counterfactuals. Although ubiquitous, counterfactual statements are
anything if not tricky (and often, here by example, ambiguous).

Of course, if no signalling theorems were false we could signal to each other using
\emph{real} frequencies (rather than \emph{relative} frequencies).  One could use,
say, $10,000$ of these experiments and work out the frequencies predicted by each
such that $S'_1$ makes the same choice of parameter on all $10,000$ sub-experiments;
either she measures the $x$-spin on all the sub-experiments or on none.  This is even
more effort to go to in order to transmit a bit of information, but it could be done
in principle; this is why we are glad no-signalling theorems are true. But still,
$S'_1$ and $S'_2$ have collaborated significantly which might\----we have not ruled
it out\----be the logical equivalent of a common cause.

So, we have the option to call the relative frequency of an event, given a fixed
ensemble of results, an `objective' relative frequency (or perhaps a propensity).
When choices are being made, however, we do not even have a well-defined experiment.
In order to make it well-defined we must fix the prior-frequency so as to fix the
distinct super-ensemble we are discussing. In order to make relative frequencies
`objective' we must first know the prior-frequencies that are implicit in their
definition. Why call something `objective' at a single trial if you must first
restrict it further than you can legitimately at a single trial? For an experiment
at spacelike separation this notion of prior-frequency obviously introduces a
fundamental ambiguity that we cannot remove. To summarise, in order to prove
causality one would need to be able to define a notion of probability that is
well-defined at a single trial and within a small spacetime region. We have shown
that relative frequency (and thus also propensity) interpretations do not satisfy
these requirements. Thus the only options we have left, bar particularly exotic
examples, are subjective notions of relative frequencies (not well-defined at a
single trial) or Bayesian probabilities (at least sometimes well-defined at a single
trial).

This notion of prior-frequency is conceptually very similar to the Bayesian notion of
prior-probability, and, ironically, frequentists often use this notion to try and
attack the Bayesian viewpoint with claims of circularity. The same claim of
circularity can be applied to the frequentist view. This analysis would suggest that
frequentists and Bayesians are more similar than either might want to accept.
Bayesian probability is subjective and is a form of logical inference that observers
make. From this point-of-view the `real' probability is a misnomer, but the
probabilities predicted by each observer are apt. If we relativise probabilities to
being defined only with respect to a certain observer then we trivially do not come
across the problems discussed above. The problems only arise when we try to define a
probability objectively in each run of the experiment. This cannot be done using
relative frequencies due to the inherent ambiguity expressed by prior-frequencies.
It cannot be done for Bayesian probability since such a global probability
assignment goes against the very definition of probability in this framework. As we
discussed above, the probabilities inferred solely by each observer can be made
passively Lorentz invariant, and in this case the probabilistic compatibility of SR
and SQT is made manifest.

\subsection*{On the Relations to Other Work}

The above argument is similar in spirit to the non-relativistic
situation discussed by Anastopoulos$^{(\ref{Anast04})}$. He notes
that relative frequencies are \emph{necessarily} additive, and
if, by presumption, the actual results of experiments are to be
given by relative frequencies then we have a conflict between
these two notions due to the non-additivity of quantum
frequencies. We have two choices; either we accept relative
frequencies and accept that the frequencies depend upon the
physical sampling used in an experiment, or we make drastic
changes to our notion of relative frequency. This argument also links
in with the Consistent Histories (CH) programme$^{(\ref{Grif84},
\ref{Omne88},\ref{GH90}, \ref{Isham94})}$ where only additive probabilities are discussed.
Thus, CH is not in conflict with the notion of relative
frequency; although rarely is the relative frequency of elements
in ensembles actually discussed in papers on CH.

A very interesting paper by Aerts$^{(\ref{Aerts02})}$ has recently
suggested a new form of probability calculus that takes account of
the different limits of relative frequency (due to the change in
observer, or context). If one wishes to use a relative frequency
interpretation then such a novel notion of frequency would seem apt;
however, it is not clear exactly how this `subset' frequency is to be
interpreted in actual experiments. Khrennikov also discusses Bell
inequalities while taking such delicate issues into
account$^{(\ref{Khrennik02})}$.

In this paper we suggest that relative frequencies
must be calculated ``relative'' to the sampling used by
observers, and that any other notion is nonsensical. This,
however, is slightly different to the way Rovelli uses the term
in his inspirational paper on relational quantum
mechanics$^{(\ref{Rovel96})}$. Here, we have not discussed the
notion of observers which measure upon whole entangled systems
(\emph{i.e.} the next ontological level up; someone who measures
the measurements of the subsystem observers). This is
because it is not clear how to interpret such an observer in
regards to spacelike separated subsystems; such an observer
would necessarily be `nonlocal'. This may, however, be a misnomer
in the sense that nonlocal observers are easy to imagine and
might be used pedagogically.

\subsection*{Conclusion}

We conclude that the Heisenberg picture should always be used in discussing spacelike
separated events because any conflict between SR and SQT seems to be manifest
through the Schr\"{o}dinger picture. Within the Heisenberg picture\----as long as we
use relative frequencies and are willing to discuss pre- and post-selection of
ensembles\----then the interpretation is consistent with changes in time-ordering.
We could also use Bayesian probabilities consistently also; we have only used
relative frequencies here for pedagogical reasons. We would like to emphasise again
the subtle distinction between the notions of causal and logical connections between
events. These notions are, of course, interlinked\----perhaps even
complimentary\----, and it is very difficult to differentiate them cogently. In
order to invoke causal relations one must discuss the choices between
counterfactually distinct cases. But when invoking relative frequencies it turns out
that each choice of a choice is invoked using a distinct super-ensemble. This blocks
any inference of causal relations when just invoking relative frequencies.

As regards the three main notions of probability, we conclude that:

\begin{enumerate}

\item Relative frequencies are properties of ensembles and, if one wishes for consistency,
one must always discuss the limits of such frequencies with respect to the ensemble or
super-ensemble used. If one wishes to use relative frequency then one must note that this
notion of ``the very definition of relative frequencies depend upon the prior-frequency of
measurements'' is a \emph{tautology}; it simply must be taken into account.

\item Bayesian probabilities are subjective (but, remember, not wholly
subjective\footnote{Nor, in some interpretations, uniquely defined for given exhaustive
prior-information.}) representations of an observer's knowledge. Prior-probability is
taken into account naturally within this view, and we argue that this is the most cogent
way to discuss probabilities because, although it is a lot more complicated than
discussing relative frequencies, it is foundationally stronger. Also, there is no problem
of using infinite trials in the Bayesian framework. A recent, but instantly classic, text
on Bayesian inference is the Gospel of Jaynes$^{(\ref{JaynesBOOK})}$. Any issues with
causality become nullified within this view as probabilities are updated by inductive
inference (rather than updated objectively).

\item `Objective' probabilities have been treated as a misnomer
within this paper exactly because the relative frequencies of
elements are taken to be logically different for different prior-frequencies of
measurements. If there is to be a notion of `objective'
probabilities (\emph{i.e.} propensities) then the `objective'
prior-frequency of measurements must be taken into account.

\end{enumerate}

In order to invoke any tentative `inconsistency' between SR and SQT one has to invoke
further assumptions (like in Bell's inequalities).  SQT is a theory of probabilities so
one must analyse the compatibility of SQT with SR only on a probabilistic level; and in
this case SR and SQT do not peacefully coexist\----they are completely compatible.  It is
rather observers (and their inferences) that peacefully coexist.

\subsection*{Acknowledgements}

I would like to acknowledge EPSRC for funding this work, and I would like to thank George
Jaroszkiewicz for generous comments on pedagogy. I would also like to thank two anonymous
referees for very constructive comments that helped structure this paper in a far more
coherent way.

\subsection*{References}

\begin{enumerate}

\item \label{Shimon83} A. Shimony, ``Controllable and uncontrollable
non-locality'' in \emph{Foundations of Quantum Mechanics in the Light of
New Technology}, S. Kamefuchi ed. (Phys. Soc. Japan, Tokyo, 1983)

\item \label{Peres00} A. Peres, ``Classical interventions in quantum
systems. II. Relativistic invariance'' \emph{Physical Review A}
\textbf{61} (2000) 022117 quant-ph/9906034

\item \label{Stapp03} H. P. Stapp, ``A Bell-type theorem without hidden
variables'' \emph{Am. J. Phys.} \textbf{72} (2004) 30-33

\item \label{Shimon04} A. Shimony, ``An Analysis of Stapp's "A Bell-type
theorem without hidden variables"'' (2004) quant-ph/0404121
 
\item \label{MisesBOOK} R. von Mises, \emph{Probability, Statistics and Truth}
(Dover, New York, 1981)

\item \label{JaynesBOOK} E. T. Jaynes, \emph{Probability Theory: The Logic
of Science} (Cambridge University Press, 2003)

\item \label{FinettiBOOK} B. de Finetti, \emph{Theory of
Probability: A Critical Introductory Treatment} (London,
Wiley, 1974)

\item \label{Fuchs02} C. A. Fuchs, ``Quantum Mechanics as Quantum
Information (and only a little bit more)'' \emph{} \textbf{}
(2002) quant-ph/0205039

\item \label{PopperBOOK} K. P. Popper, \emph{Quantum Theory and the
Schism in Physics} (Hutchinson and Co., 1982)

\item \label{Rovel99} C. Rovelli, ``Quantum spacetime: what do we
know?'' in \emph{Physics Meets Philosophy at the Planck Scale},
C Callender and N Huggett eds. (Cambridge University Press, 2001)
gr-qc/9903045

\item \label{WeinbergBOOK} S. Weinberg, \emph{The Quantum Theory of
Felds Vol 1.} (Cambridge University Press, 1995)

\item \label{Myrvold02} W. C. Myrvold, ``On peaceful coexistence: is
the collapse postulate incompatible with relativty?''
\emph{Stud. in History and Phil. of Mod. Phys} \textbf{33}
(2002) 435-466

\item \label{HK70} K. E. Hellwig and K. Kraus, ``Formal Description
of Measurements in Local Quantum Field Theory'' \emph{Phys. Rev. D}
\textbf{1} (1970) 566-571

\item \label{Vaidman98} L. Vaidman, ``Time Symmetrized Counterfactuals
in Quantum Theory'' \emph{Found.Phys.} \textbf{29} (1999) 755-765
quant-ph/9807075

\item \label{BellBOOK} J. S. Bell, ``Speakable and unspeakable in
quantum mechanics'' (Cambridge University Press, 1987)

\item \label{Anast04} C. Anastopoulos, ``On the relation between
quantum mechanical probabilities and event frequencies'' Annals
of Physics \textbf{313} (2004) 368- quant-ph/0403207

\item \label{Grif84} R. B. Griffiths, ``Consistent Histories and the
Interpretation of Quantum Mechanics'' \emph{Jour. Stat. Phys}
\textbf{36} (1984) 219-273

\item \label{Omne88} R. Omn\'es, ``Logical reformulation of quantum
mechanics. I. Foundations'' \emph{Jour. Stat. Phys.} \textbf{53}
(1988) 933-955

\item \label{GH90} M. Gell-Mann and J. Hartle, ``Quantum Mechanics in
the light of quantum cosmology'' in \emph{Proceedings of the
Third International Symposium on the Foundations of Quantum
Mechanics in the Light of New Technology}, (Physical Society of
Japan, Tokyo, Japan, 1990), 321-343

\item \label{Isham94} C. J. Isham, ``Quantum logic and the histories
approach to quantum theory'' \emph{Jour. Math. Phys.}
\textbf{35} (1994) 2157-2185
gr-qc/9308006

\item \label{Aerts02} D. Aerts, ``Reality and Probability:
Introducing a new Type of Probability Calculus'' in
\emph{Probing the Structure of Quantum Mechanics: Nonlocality,
Computation and Axiomatics}, D. Aerts, M. Czachor and T.
Durt eds. (World Scientific, Singapore, 2002) quant-ph/0205165

\item \label{Khrennik02} A. Khrennikov, ``Frequency Analysis of the EPR-Bell
Argumentation'' \emph{Found. Phys.} \textbf{32} (2002) 1159-1174

\item \label{Rovel96} C. Rovelli, ``Relational Quantum Mechanics''
\emph{Int. Jour. Theo. Phys.} \textbf{35} (1996) 1637-1678
quant-ph/9609002

\end{enumerate}

\end{document}